\documentclass[%
reprint,showpacs,preprintnumbers,
amsmath,amssymb,aps%
]{revtex4-1}

\usepackage{graphicx}
\usepackage{dcolumn}
\usepackage{bm}

\begin{document}

\preprint{}

\title{Gaugino masses from gravitino at one loop}

\author{Jae Yong Lee}
  \email{littlehiggs@gmail.com}
\affiliation{%
Department of Physics, Korea University,\\
Seoul 136-701, Korea
}%

\date{\today}

\begin{abstract}
The N=1 supergravity coupled to chiral and Yang-Mills superfields is regarded as the ultraviolet theory of the supersymmetric Standard Model.
The gravitino acquires a mass from the hidden sector and participate in mediating supersymmetry breaking to the visible sector.
We consider the quantum effects of the gravitino and evaluate its contribution to the gaugino masses at one loop level,
which turns out to be irrespective of the SM gauge couplings.
Due to the universality of gravitino, its contribution to the gaugino masses can significantly modify
the gaugino mass relations of other supersymmetry breaking mechanisms depending on the gravitino mass.
For a given gaugino mass the gravitino mass cannot be arbitrarily large even when other supersymmetry breaking mechanisms are dominant.
\end{abstract}

\maketitle

\section{Introduction\label{sec:level1}}

As the LHC experiments continue to pile up the data
the kudos for the low-energy supersymmetry is fading away.
So it is the appropriate time to look for excuses for failure to find the low-energy supersymmetry at the LHC.
The supersymmetric partners of the Standard Model (SM) can acquire masses
from various mediating mechanisms for supersymmetry breaking in the hidden sector.
One of them is ``the gravitino'', which is the supersymmetric partner of the graviton in supergravity.
Our interest lies in the gravitino-loop contribution to the masses of the supersymmetric partners, which can never be turned off.
Fayet had already indicated that the gauginos could acquire a mass from the gravitino at quantum level
using the argument of the continuous $R$-symmetry breaking~\cite{Fayet:1977vd,Fayet:1978qc}.
In this work, we rigoruously consider the gravitino contribution to the gaugino masses at one loop level.

The gravitino is the gauge field corresponding to a local supersymmetry transformation
of supergravity~\cite{Freedman:1976xh,Deser:1976eh}.
The quantization of supergravity was studied in~\cite{Das:1976ct,Freedman:1976py,Deser:1977nt,Fung:1980fq}.
The gravitino propagator possesses not only the spin-$\frac{3}{2}$ projection component but also the spin-$\frac{1}{2}$ projection component.
The gravitino gets massive by acquiring the degrees of freedom appropriate to finite mass from the hidden sector,
which is so-called ``the supersymmetric Higgs mechanism''~\cite{Deser:1977uq}.
The quantum properties of the massive gravitino were rigorously analyzed in~\cite{Baulieu:1985wa}.
Just as the quantization procedure of Yang-Mills gauge theories with spontaneously broken symmetry involves
propagators of the gauge fields, Higgs fields and ghosts
the quantization procedure of supergravity coupled to matters with spontaneously broken supersymmetry does
those of the gravitino, goldstino and ghosts.
Thus the quantum treatments of gravitino in general necessitate not only the gravitino propagator
but also the goldstino and ghosts propagators.

The classical lagrangian for supergravity coupled to chiral and Yang-Mills matter
is given in~\cite{Cremmer:1978hn,Cremmer:1982wb,Wess:1992cp,Binetruy:2000zx}
while the quantum effective lagrangian at one loop is given in~\cite{Gaillard:1993es,Gaillard:1996ms,Choi:1997de}.
We choose the unitary gauge for the local supersymmetry, in which masses of the goldstino and ghosts are infinite.
It implies that $(i)$ both the goldstino and ghost propagators vanish
so the contribution to the gaugino masses from these spurious spin-$\frac{1}{2}$ particles
is obviously absent at one loop level,
and $(ii)$ the spin-$\frac{1}{2}$ projection part of the gravitino propagator decouples.
They simplify the evaluation of the gaugino masses at one loop level.

\section{Supergravity and the $\alpha_s$ gauge}
The gravitino is a spin-$\frac{3}{2}$ fermion, obeying the Rarita-Schwinger equation.
It is the supersymmetric partner of the graviton which is described by general relativity.
The Einstein-Hilbert lagrangian in general relativity is given as
\begin{equation}
\mathcal{L}_{\rm EH}=-\frac{1}{2\kappa^2}{\cal R},
\end{equation}
where ${\cal R}$ is the Ricci scalar and $\kappa^{-2}=(8\pi G)^{-1} =m_{\rm P}^2$ with $m_{\rm P}$
being the reduced Planck mass.
The lagrangian has a single dimensionful parameter, ``the Planck mass'', 
and can be regarded as an effective lagrangian valid up to the Planck scale.

Applying the idea of an effective theory to supergravity in the same manner, 
one can write the complete quadratic lagrangian for the gravitino, including mass and gauge fixing terms, is given by
\begin{eqnarray}\label{kinetic:g}
\mathcal{L}_{3/2}&&=\frac{1}{\kappa^2}\Big[\frac{i}{2}\varepsilon^{\mu\nu\rho\sigma}\, ^\ast\Psi_\mu\gamma_5\gamma_\nu\partial_\rho \Psi_\sigma
+im_{3/2}\, ^\ast\Psi_\mu\Sigma^{\mu\nu}\Psi_\nu\nonumber\\ 
&&\qquad +\frac{i}{2}\alpha_s\, ^\ast\Psi \cdot\gamma\,\partial\!\!\!\slash\,\gamma\cdot\Psi\Big].
\end{eqnarray}
Here the mass dimension of $\Psi_\mu$ is $\frac{1}{2}$ rather than $\frac{3}{2}$,
and Dirac conjugation is denoted by $^\ast\Psi_\mu=\Psi_\mu^\dagger\gamma^0$.
Instead one can persist in $\frac{3}{2}$ for the mass dimension of $\Psi_\mu$ with
the Planck mass entering into the coefficients of the interaction terms.
The first term is the kinetic one for Rarita-Schwinger spin-$\frac{3}{2}$ field, the second the mass term,
and the third the gauge-fixing term which embraces
the two spin-$\frac{1}{2}$ spurious modes $\gamma^\mu\psi_\mu$ and $\partial^\mu\psi_\mu$.
The gravitino mass arises from supersymmetry breaking in the hidden sector. 
Our conventions for $\gamma$-matrices are written in Appendix~\ref{appendix:1}.

The gravitino propagator is then expressed, in the $\alpha_s$ gauge, by~\cite{Baulieu:1985wa}:
\begin{eqnarray}\label{eq:gravitino1}
\frac{i}{\kappa^2} \langle \Psi_\mu\,\!  ^\ast\Psi_\nu \rangle &&= 
P^{3/2}_{\mu\nu} \frac{1}{m_{3/2}+\partial\!\!\!\slash}\nonumber\\
&&+\sum_{i,j} (P^{1/2}_{ij})_{\mu\nu}\big[\alpha_{ij}(\square)\partial\!\!\!\slash+\beta_{ij}(\square)\big],
\end{eqnarray}
where $P^{3/2}$ and $P^{1/2}$ are, respectively, spin-$\frac{3}{2}$ and spin-$\frac{1}{2}$ projection operators~\cite{Fung:1980fq}:
\begin{align}
P^{3/2}_{\mu\nu} &= \theta_{\mu\nu}-\frac{1}{3}\hat\gamma_\mu\hat\gamma_\nu,\nonumber\\
(P^{1/2}_{11})_{\mu\nu}&= \frac{1}{3}\hat\gamma_\mu\hat\gamma_\nu, \qquad \quad
(P^{1/2}_{12})_{\mu\nu}= \sqrt{\frac{1}{3}}\hat\gamma_\mu\omega_\nu, \\
(P^{1/2}_{21})_{\mu\nu}&= \sqrt{\frac{1}{3}}\omega_\mu\hat\gamma_\nu, \qquad
(P^{1/2}_{22})_{\mu\nu}= \frac{1}{3}\omega_\mu\omega_\nu\nonumber,
\end{align}
with
\begin{equation}
\omega_\mu\equiv\frac{\partial_\mu\partial\!\!\!\slash}{\square},
\quad \hat\gamma_\mu\equiv\gamma_\mu-\omega_\mu,
\quad \theta_{\mu\nu}\equiv\eta_{\mu\nu}-\omega_\mu\omega_\nu,
\end{equation}
and $\alpha_{ij}$ and $\beta_{ij}$ are functions of the box operator solutions of the matrix equations:
\begin{equation}\label{eq:abab}
\left\{
\begin{aligned}
&\alpha\cdot \tilde a\,\square + \beta\cdot b=\mathbf{1}_2,\\
&\alpha\cdot\tilde b+\beta\cdot a=0,
\end{aligned}
\right.
\end{equation}
with
\begin{align}
a&=\left(\begin{array}{cc} -2\big(1-\frac{3}{2}\alpha_s\big)& -\sqrt{3}\alpha_s \\
\sqrt{3}\alpha_s & -\alpha_s \end{array}\right),\nonumber\\
\tilde a&= \left(\begin{array}{cc} -2\big(1-\frac{3}{2}\alpha_s\big)& \sqrt{3}\alpha_s \\
-\sqrt{3}\alpha_s & -\alpha_s \end{array}\right),\\
b&=m_{3/2}\left(\begin{array}{cc} -2 & -\sqrt{3} \\ -\sqrt{3} & 0 \end{array}\right), \quad
\tilde b=m_{3/2}\left(\begin{array}{cc} -2 & \sqrt{3} \\ \sqrt{3} & 0 \end{array}\right).\nonumber
\end{align}
The spin-$\frac{3}{2}$ projection part has a single pole at $k^2=-m^2_{3/2}$
while the spin-$\frac{1}{2}$ projection part does in general two different poles.
This analysis is analogous to that of the Yang-Mills gauge theory with spontaneously broken symmetry
so that the $\alpha_s$ gauge in supergravity corresponds to the $R_\xi$ gauge in the Yang-Mills theory. 

The gravitino propagator can be decomposed into two parts: one is chirality-preserving
while the other chirality-flipping.
Because the gravitino mass appears in the numerator of the chirality-flipping part
the two-component spinor techniques~\cite{Binetruy:2000zx,Dreiner:2008tw}
is an efficient method for our purpose.
Note that the gauge dependence of the chirality-flipping propagator shows up only in spin-$\frac{1}{2}$ projection part.
 
Using two-component spinor techniques, the gravitino propagator in (\ref{eq:gravitino1}) is decomposed into four parts.
Among them we are interested in the upper $2\times 2$ component for chirality-flipping propagator.
The corresponding spin projection operators are given by
\begin{align}
P^{3/2}_{\mu\nu} &= \eta_{\mu\nu}-\frac{1}{3}\Big(\sigma_\mu\bar\sigma_\nu-\frac{k_\nu\sigma_\mu k \cdot \bar\sigma}{k^2}
-\frac{k_\mu k\cdot\sigma\bar\sigma_\nu}{k^2}\nonumber\\
&\qquad+4\frac{k_\mu k_\nu}{k^2}\Big),\nonumber\\
P^{1/2}_{11,\mu\nu}&=\frac{1}{3}\Big(\sigma_\mu\bar\sigma_\nu-\frac{k_\nu\sigma_\mu k \cdot \bar\sigma}{k^2}
-\frac{k_\mu k\cdot\sigma\bar\sigma_\nu}{k^2}+\frac{k_\mu k_\nu}{k^2}\Big),\nonumber\\
P^{1/2}_{12,\mu\nu}&= \frac{1}{\sqrt{3}}\Big(\frac{k_\nu \sigma_\mu k\cdot\bar\sigma}{k^2}-\frac{k_\mu k_\nu}{k^2} \Big),\\
P^{1/2}_{21,\mu\nu}&= \frac{1}{\sqrt{3}}\Big(\frac{k_\mu k\cdot\sigma \bar\sigma_\nu}{k^2} -\frac{k_\mu k_\nu}{k^2}\Big),\nonumber\\
P^{1/2}_{22,\mu\nu}&=\frac{k_\mu k_\nu}{k^2},\nonumber
\end{align}
where $k^\mu$ is the gravitino momentum.
Then the chirality-flipping propagators both for spin-$\frac{3}{2}$ and spin-$\frac{1}{2}$ projection are respectively given by
\begin{widetext}
\begin{equation}
\langle \psi_\mu\overline{\psi}_\nu \rangle_\frac{3}{2} =
(-i) \kappa^2\Big[ \eta_{\mu\nu}-\frac{1}{3}\Big(\sigma_\mu\bar\sigma_\nu-\frac{k_\nu\sigma_\mu k \cdot \bar\sigma}{k^2}
-\frac{k_\mu k\cdot\sigma\bar\sigma_\nu}{k^2}+4\frac{k_\mu k_\nu}{k^2}\Big)\Big] \frac{m_{3/2}}{k^2+m^2_{3/2}}
\end{equation}
\end{widetext}
and
\begin{widetext}
\begin{equation} 
\begin{aligned}
\langle \psi_\mu\overline{\psi}_\nu \rangle_\frac{1}{2} =
(-i)\kappa^2 &\Big[\,\frac{1}{3}\Big(\sigma_\mu\bar\sigma_\nu-\frac{k_\nu\sigma_\mu k \cdot \bar\sigma}{k^2}
-\frac{k_\mu k\cdot\sigma\bar\sigma_\nu}{k^2}+\frac{k_\mu k_\nu}{k^2}\Big)\beta_{11}
+\frac{k_\mu k_\nu}{k^2} \beta_{22}\\
&+ \frac{1}{\sqrt{3}}\Big(\frac{k_\nu \sigma_\mu k\cdot\bar\sigma}{k^2}-\frac{k_\mu k_\nu}{k^2} \Big)\beta_{12}
+ \frac{1}{\sqrt{3}}\Big(\frac{k_\mu k\cdot\sigma \bar\sigma_\nu}{k^2} -\frac{k_\mu k_\nu}{k^2}\Big) \beta_{21}\Big],
\end{aligned}
\end{equation}
\end{widetext}
where $\beta_{ij}$ is the solution of Eq.~(\ref{eq:abab}).
Our conventions for $\sigma$-matrices are written in Appendix~\ref{appendix:1}.

\begin{figure}[tbp]
\centering
\includegraphics[width=0.25\textwidth]{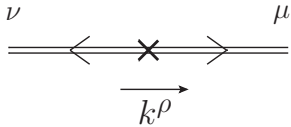}
\hfill
\caption{\label{fig:1} The chirality-flipping gravitino propagator. The cross stands for mass insertion.  }
\end{figure}

The gauge dependence of the chirality-flipping propagator is realized
by the value of $\alpha_s$ through $\beta$.
In this work, we take $a_s=0$, where $\beta$ is given by
\begin{equation}
\beta=\left(\begin{array}{cc} 0 & -\frac{1}{\sqrt{3}}\frac{1}{m_{3/2}} \\ 
-\frac{1}{\sqrt{3}}\frac{1}{m_{3/2}} & \frac{2}{3}\frac{1}{m_{3/2}} \end{array}\right).
\end{equation}
Note that $\beta$ does not depend on the momentum so that the spin-$\frac{1}{2}$ projection component decouples.

\section{Feynman rules in the unitary gauge}
The quantum lagrangian contains the goldstino and ghost fields corresponding to the gravitino.
The quadratic terms of the two ``spurious'' spin-$\frac{1}{2}$ modes are given by Eq.~(4.33) in \cite{Baulieu:1985wa}
and their mass matrix is given by
\begin{equation}
M^2_{1/2}=\frac{m^2_{3/2}}{\alpha_s^2}\left[\begin{array} {cc} 
\alpha_s\big(\alpha_s-\frac{3}{2}\big) & -\sqrt{3} \alpha^2_s \\
\sqrt{3}\alpha_s(2-\alpha_s) & 3\alpha_s\big(\alpha_s-\frac{1}{2}\big)\end{array}\right],
\end{equation}
with the mass eigenvalues 
\begin{equation}
m_{3/2}\pm m_{3/2} \sqrt{1-\frac{3}{2\alpha_s}}.
\end{equation}
Note that there is a typo at the $(2,1)$ element in the mass matrix given by the original paper~\cite{Baulieu:1985wa}.
In the limit of $\alpha_s\to 0$, both the goldstino and ghost fields get infinite masses and decouple.
This is  ``the unitary gauge''. 
 
In the unitary gauge of spontaneously broken Yang-Mills theories,
there exists an extra quadratically divergent Higgs self-interaction term~\cite{GrosseKnetter:1992nn}.
Likewise one expects an extra quadratically divergent self-interaction term for fields
which belong to the same multiplet with the goldstino in supergravity.
However it is irrelevant to the visible sector and has nothing to do with the gravitino contribution to the gaugino mass.
So it is safe to deal with the interaction terms from the classical supergravity lagrangian,
which is given by (4.5.29) in~\cite{Binetruy:2000zx}. 
We take the minimal K\"{a}hler potential for the chiral multiplets in flat spacetime limit.

\begin{figure}[tbp]
\centering
\includegraphics[width=0.5\textwidth]{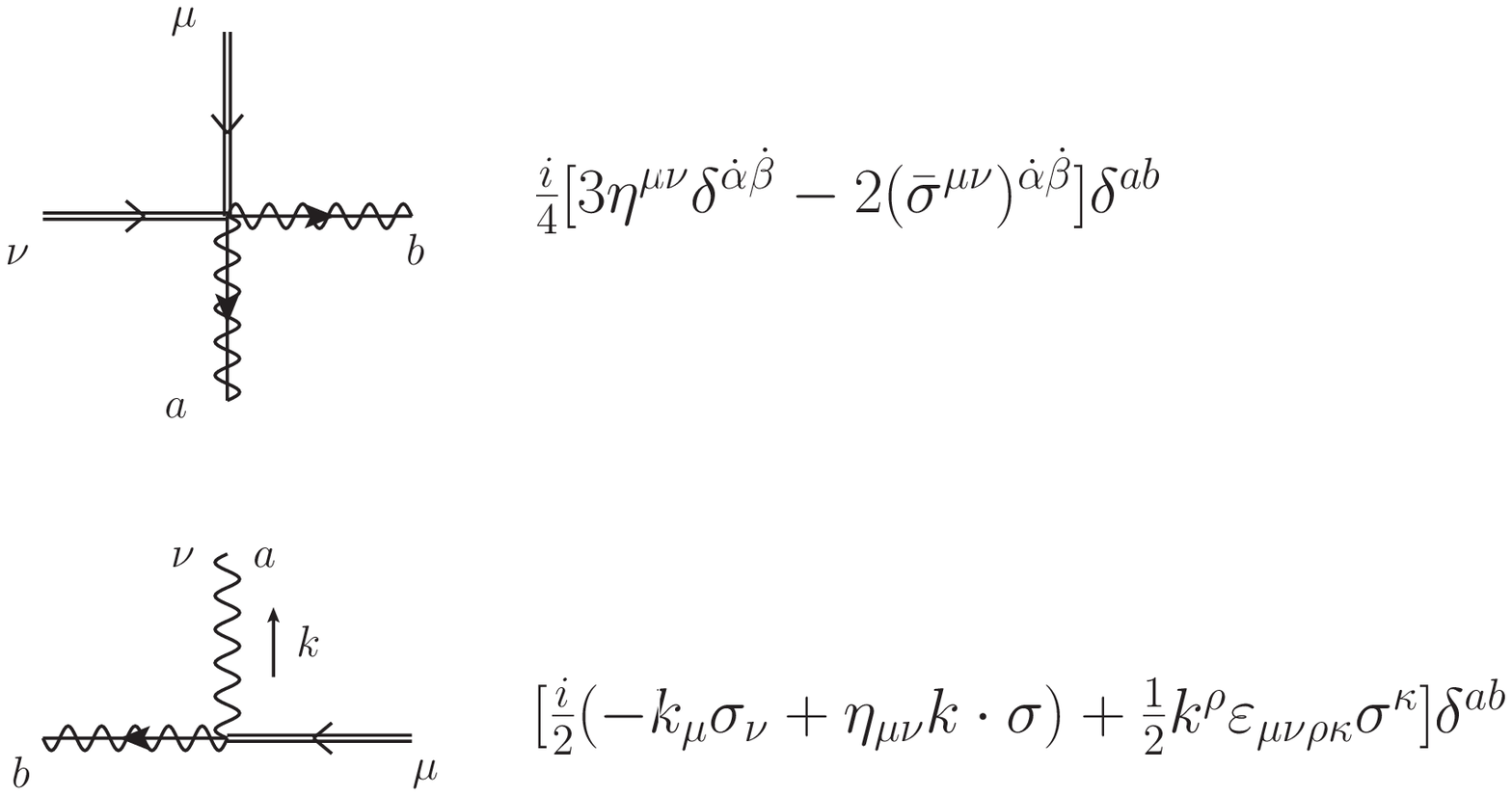}
\caption{\label{fig:2} The Feynman rules for the interactions in (\ref{eq:lag1}).
The double lines, dashed lines and wiggle lines denote gravitino, Yang-Mills gauge and gaugino fields, respectively.}
\end{figure}

We do not write down the complete lagrangian here.
Instead, we put down the gaugino and gravitino interaction terms relevant to our purpose:
\begin{equation}\label{eq:lag1}
\begin{aligned}
{\cal L}_1 &=\frac{1}{16}[\lambda^{a}\lambda^{a} \bar\psi_\mu(3g^{\mu\nu}-2\bar\sigma^{\mu\nu})\bar\psi_\nu\\
&\qquad+\bar\lambda^{a}\bar\lambda^{a} \psi_\mu(3g^{\mu\nu}-2\sigma^{\mu\nu})\psi_\nu],\\
{\cal L}_2 &=  \frac{i}{2} [\partial^\mu {\cal A}^{\nu c} -\partial^{\nu}{\cal A}^{\mu c} ]
\big[(\psi_\mu \sigma_\nu \bar\lambda^{c}-\bar\psi_\mu\bar\sigma_\nu\lambda^{c})\\
&\qquad-\frac{i}{2}\varepsilon_{\mu\nu\rho\kappa}(\psi^\rho \sigma^\kappa \bar\lambda^{c}+\bar\psi^\rho\bar\sigma^\kappa\lambda^{c})\big],\\
\end{aligned}
\end{equation}
where $\lambda^a, {\cal A}^{\mu a}, \psi_\mu,$ are gaugino, Yang-Mills gauge, gravitino fields, respectively.
Note that no gauge coupling appears  in ${\cal L}_{1,2}$.
The relevant Feynman rules are given in Figure \ref{fig:2}.

\section{Gaugino masses from gravitino}
\begin{figure}[tbp]
\centering 
\includegraphics[width=0.5\textwidth]{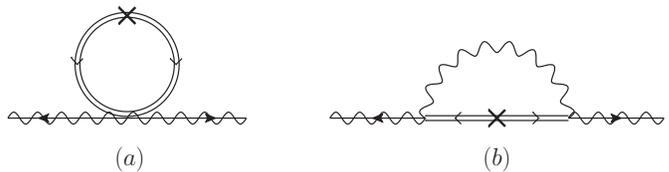}
\hfill
\caption{\label{fig:3} Gravitino loop contribution to gaugino mass.}
\end{figure}

The Feynman diagrams for the gaugino mass from gravitino loops are shown in Figure~\ref{fig:3}.
One may expect that the one-loop diagrams lead to divergences
which can be removed by introducing counterterms.
However, supergravity lagrangian itself has no gaugino mass term,
which is forbidden by the Yang-Mills gauge symmetry.
So dimensional regularization scheme cannot be used to evaluate the one-loop diagrams.
Instead we use an ultraviolet cutoff regularization,
in which all momentum integrals are limited naturally to a region $q^2 < \Lambda^2$, with $\Lambda$ being an ultraviolet cutoff.
Moreover we surmise in Section 2 that supergravity is an effective theory up to the Planck scale
which is nothing but an ultraviolet cutoff.
As a result, no ultraviolet divergences can occur.
For instance, using an ultraviolet cutoff regularization one gets the one-loop self-energy in $\phi^4$ theory as follows
\begin{equation}\label{eq:div}
\int^\Lambda \frac{d^4q}{(2\pi)^4} \frac{1}{q^2+m^2} = 
\frac{i}{16\pi^2}\Big[\Lambda^2-m^2\log\Big(\frac{\Lambda^2}{m^2}+1\Big)\Big].
\end{equation}

Now we evaluate the Feynman diagrams in Figure~\ref{fig:3} using an ultraviolet cutoff regularization
and obtain the gaugino mass at one loop level as
\begin{equation}\label{eq:mhalfmass}
m_{1/2} = \frac{m_{3/2}}{16\pi^2}\Big[\frac{\Lambda^2}{m_P^2}-\frac{m^2_{3/2}}{m^2_{\rm P}}\log\Big(\frac{\Lambda^2}{m^2_{3/2}}+1\Big)\Big],
\end{equation}
which does not include the Yang-Mills gauge coupling,
implying that different gauginos acquire the same mass from gravitino at one loop level.
Furthermore, the scale dependence of the gaugino masses stems only from that of the gravitino mass. 
However, the universal feature cannot survive at higher loop level 
because gauge couplings in the gaugino masses begin to appear at two loop level.
Note that spin-$\frac{1}{2}$ projection part of the gravitino propagator
does not contribute to the gaugino mass.

Just as one cancels fermonic loop contributions to the SM Higgs mass by adding bosonic loop contributions
one may try to cancel the $\Lambda^2$ in the first term of Eq.~(\ref{eq:mhalfmass})
by introducing a supersymmetric counterpart which involves a graviton loop.
But there are no such diagrams because both spin statistics and Yang-Mills gauge symmetry do not allow for them.
Unlike the hierarchy problem of the SM Higgs mass the $\Lambda^2$ dependence in the gaugino mass does not cause
a hierarchy problem due to the Planck mass suppression.
For $m_{3/2}\ll m_P$, the ratio of gaugino mass to gravitino mass is given by
\begin{equation}
\frac{m_{1/2}}{m_{3/2}} \sim \frac{1}{16\pi^2} \frac{\Lambda^2}{m_P^2},
\end{equation}
implying that the gravitino mass cannot be arbitrarily large for a given gaugino mass.
This holds true even when the contribution to the gaugino mass from other supersymmetry breaking mechanisms is dominant.
Furthermore, with a naive expectation $\Lambda = m_P$ one gets
\begin{equation}
\frac{m_{1/2}}{m_{3/2}} \sim \frac{1}{16\pi^2}.
\end{equation}

It is well known that performing an ultraviolet cutoff regularization procedure in general
leads to unwanted explicit gauge-violating contributions. 
However there is a remedy for it~\cite{Polchinski:1983gv,Kizilersu:2000dk,Gu:2006zw}:
imposing translational invariance in momentum space on to an ultravioet cutoff regularization.
In our case it is achieved by cleverly choosing the gauges: 
the unitary gauge for local supersymmetry transformation and the Feynman gauge for the Yang-Mills gauge symmetry, respectively.
To be specific, the dependence of the external momentum in Figure~\ref{fig:3} (b) disappears at the end
so one can maintain the translational invariance.

\section{Summary and outlook}
We considered the gaugino mass by quantum gravitino mediation.
It is found that the gaugino masses at one loop level are the same irrespective of the kinds of Yang-Mills gauge symmetries.
Once the contribution is comparable to those of other supersymmetry breaking mechanisms
it will significantly change the gaugino mass spectrums.
All depend on the gravitino mass and the momentum UV cutoff.
For instance, the gaugino mass relations from anomaly mediation should be clearly modified
because they depend on the gauge couplings.
We will consider the gravitino-loop contributions to sfermon masses in the near future.
It is also expected that soft supersymmetry parameters of the MSSM from anomaly mediation
may be understood using the explicit one-loop effective supergravity lagrangian~\cite{Gaillard:1993es,Gaillard:1996ms}.
These may shake the footing of the CMSSM or other low-energy supersymmetry models,
providing any theoretical reasons why low-energy supersymmetry is not found at the LHC.

\begin{acknowledgments}
We would like to thank Y.W. Yoon for many valuable discussions.
This work is supported in part by Basic Science Research Program through the National Research Foundation of Korea(NRF)
funded by the Ministry of Education, Science and Technology(2011-0003974),
by Mid-career Research Program through the NRF grant funded by the MEST(2011-0027559) and 
by NRF Research Grant 2012R1A2A1A01006053.
\end{acknowledgments}

\appendix
\section{Notations}\label{appendix:1}
In our notations, the metric tensor $\eta^{\mu\nu} (\mu,\nu=0,1,2,3)$ is defined as
\begin{equation}
\eta^{\mu\nu}= (-,+,+,+).
\end{equation}
The totally antisymmetric symbol $\varepsilon_{\mu\nu\lambda\kappa}$ is normalised such as
\begin{equation}
\varepsilon_{0123}=+1,\quad \varepsilon^{0123}=-1.
\end{equation}
We use the $\sigma$-matrices,
\begin{equation}
\begin{aligned}
&\bar\sigma_\mu = \sigma^\mu=(\mathbf{1}_2,\sigma^i),\qquad \sigma^i= \sigma_i,\\
&\bar\sigma^\mu=(-\mathbf{1}_2,\sigma^i),\qquad \sigma_\mu=(-\mathbf{1}_2,\sigma_i).
\end{aligned}
\end{equation}
 
In the Weyl basis the Dirac matrices are given by
\begin{equation}
\gamma^\mu\equiv\left(\begin{array}{ccc} 0 & &\sigma^\mu \\ \bar{\sigma}^\mu & & 0 \end{array}\right),
\end{equation}
and the traceless antisymmetric combinations are defined as
\begin{equation}
\Sigma^{\mu\nu}\equiv\frac{1}{4}(\gamma^\mu\gamma^\nu-\gamma^\nu\gamma^\mu)
=\left(\begin{array}{cc} \sigma^{\mu\nu} & 0 \\ 0 & \bar{\sigma}^{\mu\nu}\end{array}\right),
\end{equation}
where
\begin{equation}
\sigma^{\mu\nu}\equiv\frac{1}{4}(\sigma^\mu\bar\sigma^\nu-\sigma^\nu\bar\sigma^\mu),\qquad
\bar{\sigma}^{\mu\nu}\equiv\frac{1}{4}(\bar\sigma^\mu\sigma^\nu-\bar\sigma^\nu\sigma^\mu).
\end{equation}

A Majorana spinor $\Psi_\mu$ is made of a Weyl spinor $\psi_\alpha$ with two components, $\alpha=1,2$
and its complex conjugate $\overline{\psi}^{\dot\alpha}, (\dot\alpha=\dot{1},\dot{2})$:
\begin{equation}
\Psi_\mu = \left(\begin{array}{c} {(\psi_\mu)}_\alpha \\ (\overline{\psi}_\mu)^{\dot\alpha}\end{array}\right),\quad
\overline{\Psi}_\mu= \left(\begin{array} {cc} (\psi_\mu)^\alpha, & ({\overline{\psi}_\mu)_{\dot\alpha}}\end{array}\right).
\end{equation}


\thebibliography{gravitino}

\bibitem{Fayet:1977vd}
  P.~Fayet,
  Phys.\ Lett.\ B {\bf 70} (1977) 461.
  
\bibitem{Fayet:1978qc}
  P.~Fayet,
  Phys.\ Lett.\ B {\bf 78} (1978) 417.
  
\bibitem{Freedman:1976xh}
  D.~Z.~Freedman, P.~van Nieuwenhuizen and S.~Ferrara,
  Phys.\ Rev.\ D {\bf 13} (1976) 3214.
  
\bibitem{Deser:1976eh}
  S.~Deser and B.~Zumino,
  Phys.\ Lett.\ B {\bf 62} (1976) 335.

\bibitem{Das:1976ct}
  A.~K.~Das and D.~Z.~Freedman,
  Nucl.\ Phys.\ B {\bf 114} (1976) 271.

\bibitem{Freedman:1976py}
  D.~Z.~Freedman and P.~van Nieuwenhuizen,
  Phys.\ Rev.\ D {\bf 14} (1976) 912.

\bibitem{Deser:1977nt}
  S.~Deser, J.~H.~Kay and K.~S.~Stelle,
  Phys.\ Rev.\ Lett.\  {\bf 38} (1977) 527.
  
\bibitem{Fung:1980fq}
  M.~K.~Fung, P.~Van Nieuwenhuizen and D.~R.~T.~Jones,
  Phys.\ Rev.\ D {\bf 22} (1980) 2995.
  
\bibitem{Deser:1977uq}
  S.~Deser and B.~Zumino,
  Phys.\ Rev.\ Lett.\  {\bf 38} (1977) 1433.
  
\bibitem{Baulieu:1985wa}
  L.~Baulieu, A.~Georges and S.~Ouvry,
  Nucl.\ Phys.\ B {\bf 273} (1986) 366.
  
\bibitem{Cremmer:1978hn}
  E.~Cremmer, B.~Julia, J.~Scherk, S.~Ferrara, L.~Girardello and P.~van Nieuwenhuizen,
  Nucl.\ Phys.\ B {\bf 147} (1979) 105.

\bibitem{Cremmer:1982wb} 
  E.~Cremmer, S.~Ferrara, L.~Girardello and A.~Van Proeyen,
  Phys.\ Lett.\ B {\bf 116}, 231 (1982).

\bibitem{Wess:1992cp}
  J.~Wess and J.~Bagger,
  Princeton, USA: Univ. Pr. (1992) 259 p

\bibitem{Binetruy:2000zx}
  P.~Binetruy, G.~Girardi and R.~Grimm,
  Phys.\ Rept.\  {\bf 343} (2001) 255
  [hep-th/0005225].
  
\bibitem{Gaillard:1993es}
  M.~K.~Gaillard and V.~Jain,
  Phys.\ Rev.\ D {\bf 49} (1994) 1951
  [hep-th/9308090].
  
\bibitem{Gaillard:1996ms}
  M.~K.~Gaillard, V.~Jain and K.~Saririan,
  Phys.\ Rev.\ D {\bf 55} (1997) 883
  [hep-th/9606052].
  
\bibitem{Choi:1997de}
  K.~Choi, J.~S.~Lee and C.~Munoz,
  Phys.\ Rev.\ Lett.\  {\bf 80} (1998) 3686
  [hep-ph/9709250].
  
\bibitem{Dreiner:2008tw}
  H.~K.~Dreiner, H.~E.~Haber and S.~P.~Martin,
  Phys.\ Rept.\  {\bf 494} (2010) 1
  [arXiv:0812.1594 [hep-ph]].

\bibitem{GrosseKnetter:1992nn}
  C.~Grosse-Knetter and R.~Kogerler,
  Phys.\ Rev.\ D {\bf 48} (1993) 2865
  [hep-ph/9212268].
  
\bibitem{Polchinski:1983gv}
  J.~Polchinski,
  Nucl.\ Phys.\ B {\bf 231} (1984) 269.

\bibitem{Kizilersu:2000dk}
  A.~Kizilersu, T.~Sizer and A.~G.~Williams,
  hep-ph/0001147.
  
\bibitem{Gu:2006zw}
  Y.~Gu,
  J.\ Phys.\ A {\bf 39} (2006) 13575.
\end{document}